\documentclass[twocolumn,showpacs,preprintnumbers,amsmath,amssymb]{revtex4}
\usepackage{graphicx}
\usepackage{dcolumn}
\usepackage{bm}

\begin{document}

\preprint{APS/123-QED}
\title{Long-wave instability and growth rate of the
 inviscid shear flows }
\author{Liang Sun}
\email{sunl@ustc.edu.cn; sunl@ustc.edu} \affiliation{ 1. School of
Earth and Space Sciences University of Science and Technology of
China, Hefei, 230026, P.R.China.\\
2. LASG, Institute of Atmospheric Physics, Chinese Academy of
Sciences, Beijing 100029, China}

\date{\today}
\begin{abstract}
In this paper, we studied the long-wave instability of the shear
flows. When the wavenumber of perturbation is larger than the
critical value, the flow is always neutrally stable. First, we
obtain a new upper bound for the neutral wavenumber $k_1\leq
(p^2-1)\mu_1$, where $p>1$ and $\mu_1$ is the smallest eigenvalue
of Poincar\'{e}'s problem. Second, we find a new upper bound for
the imaginary part of the complex phase velocity $c_i \leq k_1
\Delta U/\sqrt{\mu_1}$, where $\Delta U$ is the variance of the
velocity. The new bound is finite for all $k>0$ similar to the
Howard's semicircle theorem, while the previous ones by Craik and
Banerjee et al would be infinity as $k\rightarrow 0$. Third, we
find a new upper bound of growth rate $\omega_i \leq (p-1)
\sqrt{\mu_1} \Delta U$. All the new bounds are much more strict
than the previous ones by H{\o}iland, Howard, Craik and Banerjee
et al. Our results also extend the inverse energy cascade theory
by Kraichnan. As shear instability is due to long-wave
instability, it implies that the truncation of long-waves may
change the instability of shear flows.

\end{abstract}
 \maketitle

Shear instability caused by velocity shear is one of most
important factors in flow instabilities. Although the mechanism of
shear instability are yet to be fully revealed, it has been
applied to explain instability in mixing layers, jets in pipes,
wakes behind cylinders, etc. Some simple models have been employed
to study shear instability, including the Kelvin-Helmholtz (K-H)
model, piecewise linear velocity profile \cite{Rayleigh1894},
continued velocity profile $U(y)$ \cite{Rayleigh1880}, etc.
Rayleigh first proved a necessary criterion for instability, i.e.,
Inflection Point Theorem, which is also called Rayleigh-Kuo
theorem \cite[]{CriminaleBook2003} for Kuo's generalization to
barotropic geophysical flows in the $\beta$ plane
\cite[]{KuoHL1949}. According to the theorem, a necessary
condition for instability is $U''(y_s)=0$, where $y_s$ is the
inflection point and $U_s=U(y_s)$ is the velocity at $y_s$. Then,
Fjortoft found a stronger necessary criterion that $U''(U-U_s)<0$
somewhere for instability \cite{Fjortoft1950}. For some special
flow velocity profile (e.g. symmetric or monotone in $y$),
Tollmien \cite{Tollmien1935}, then von Mises and Friedrichs
\cite{Friedrichs1971} pointed out that there exist unstable
solutions if $U''_s=0$. However, Tollmien's assumptions, monotonic
or non-monotonic but symmetric, are neither sufficient nor
necessary for instability \cite{SunL2007ejp,SunL2008cpl}.

In the following studies
\cite{Friedrichs1971,Craik1972,Banerjee2000,SunL2007ejp,SunL2008cpl},
it is found that two functions are determinative. One is the
auxiliary function $f(y)=-U''/(U-U_s)$, the other is the
eigenvalue of Poincar\'{e}'s problem (see
Eq.(\ref{Eq:stable_paralleflow_Poincare}) behind)
\cite{Craik1972,Banerjee2000,SunL2007ejp}. Suppose that $\mu_1$ is
the smallest eigenvalue of Poincar\'{e}'s problem, a more strictly
sufficient criterion for stability was obtained that the flow is
stable if $0<f(y)<\mu_1$
\cite{Craik1972,Banerjee2000,SunL2007ejp}. Otherwise, the flow
might be unstable if $f(y)>\mu_1$. As the maximum of $f(y)$ is
bigger than $\mu_1$, we use $p^2\mu_1$ to present this maximum,
where $p$ is a positive number. Our previous studies have
investigated the cases when the flow is stable with $p<1$. Here we
need to consider the cases of $p>1$, where the flow is unstable.

For the unstable flows, there are two theoretical ways of studies.
The first one is the estimation of the complex velocity of the
unstable waves, which leads to the Howard's semicircle theorem
\cite{Howard1961,Drazin2004}.  It is recognized that the shear
instability is long-wave instability. The disturbances with
short-waves $k>k_0$ are always neutral stable, where $k_0$ is a
critical wavenumber subject to $k_0^2=(p^2-1)\mu_1$
\cite{Craik1972,Banerjee2000,SunL2006b}. In a less known paper,
Craik also obtained some new bounds for $k_0$ by expressing the
Rayleigh problem into a Green's function and using H\"{o}lder
inequality \cite{Craik1972}. Although the upper bounds for $k_0$
is about 10\% error, his new bounds for $c_i$ is very loose for
the small wavenumbers, even less efficient than the Howard's
semicircle theorem. The second one is the estimation of the growth
rate of the unstable waves. One can easily derive that growth
rate, $\omega_i$, must be less than or equal to half of the maxim
of vorticity, i.e., $\omega_i\leq |U'|_{\max}/2$ from the Howard's
semicircle theorem \cite{Howard1961}. This result is firstly due
to H{\o}iland \cite{Hoiland1953}. However, this estimate is too
sketchy for application purposes. For example, $U'$ is always
greater than zero even when the velocity profile has no inflection
point. Hence, this estimate is trivial for these cases. In a
recent study, Banerjee et al. improved such estimation of
$\omega_i^2<(p^2-1)|U''|^2_{\max}/(p^2\mu_1)$ by using
Rayleigh-Rize inequality \cite{Banerjee2000}.

The objective of this report is to improve the estimation of
growth rate following the previous works
\cite{Hoiland1953,Banerjee2000,SunL2006b}, but using the approach
of \cite{SunL2007ejp}. This is also a frame work of
\cite{SunL2007ejp,SunL2008cpl}.

For this purpose, long-wave instability in shear flows is
investigated via Rayleigh's equation
\cite{Rayleigh1880,Chandrasekhar1961,Huerre1998,CriminaleBook2003}.
For a parallel flow with mean velocity $U(y)$, where $y$ is the
cross-stream coordinate. The streamfunction of the disturbance
expands as series of waves (normal modes) with real wavenumber $k$
and complex frequency $\omega=\omega_r+i\omega_i$, where
$\omega_i$ relates to the grow rate of the waves. The flow is
unstable if and only if $\omega_i>0$. We study the stability of
the disturbances by investigating the growth rate of the waves,
this method is known as normal mode method. The amplitude of
waves, namely $\phi$, holds
 \begin{equation}
 (\phi''-k^2 \phi)-\frac{U''}{U-c}\phi=0,
 \label{Eq:stable_paralleflow_RayleighEq}
 \end{equation}
where $c=\omega/k=c_r+ic_i$ is the complex phase speed. The real
part of complex phase speed $c_r=\omega_r/k$ is the wave phase
speed. This equation is to be solved subject to homogeneous
boundary conditions
\begin{equation}
\phi=0 \,\, at\,\, y=-1,1.
\label{Eq:stable_paralleflow_RayleighBc}
\end{equation}

From Rayleigh's equation, we get the following equations:
\begin{equation}
\displaystyle\int_{-1}^{1}
[(|\phi'|^2+k^2|\phi|^2)+\frac{U''(U-c_r)}{|U-c|^2}|\phi|^2]\,
dy=0,
\label{Eq:stable_parallelflow_Rayleigh_Int_Rea}
 \end{equation}
and
\begin{equation}
\displaystyle c_i\int_{-1}^{1} \frac{U''}{|U-c|^2}|\phi|^2\,dy=0.
\label{Eq:stable_parallelflow_Rayleigh_Int_Img}
 \end{equation}

Before the further discussion, we need estimate the rate of
$\int_{-1}^{1} |\phi'|^2 dy$ to $\int_{-1}^{1}|\phi|^2 dy$
\cite{Mumu1993,Mumu1994,Banerjee1995,Banerjee2000,SunL2007ejp}.
This is known as Poincar\'{e}'s problem:
\begin{equation}
\mu=\frac{\int_{-1}^{1}|\phi'|^2 dy}{\int_{-1}^{1}|\phi|^2 dy},
\label{Eq:stable_paralleflow_Poincare}
\end{equation}
where the eigenvalue $\mu$ is positive definition for $\phi \neq
0$ with $\phi$ satisfies the boundary condition of
Eq.(\ref{Eq:stable_paralleflow_RayleighBc}). The smallest
eigenvalue value, namely $\mu_1$, can be estimated as
$\mu_1>\pi^2/4$ by taken $\phi_1=\cos(\pi y/2)$. As mentioned
above, an auxiliary function $f(y)=-\frac{U''}{U-U_s}$ is also
introduced for the investigations.

With the preparations above, we have such consequence. If $f(y)>0$
everywhere and maximum of $f(y)$ equals to $p^2\mu_1>\mu_1$, then
the disturbances with short-waves $k>k_0$ are always neutral
stable, where $k_0$ is a critical wavenumber subject to
$k_0^2=(p^2-1)\mu_1$ \cite{Craik1972}.

If $c_i^2\neq0$, add the product of $(U_s-c_r)/c_i$ and
Eq.(\ref{Eq:stable_parallelflow_Rayleigh_Int_Img}) to
Eq.(\ref{Eq:stable_parallelflow_Rayleigh_Int_Rea}), giving
\begin{equation}
\begin{array}{rl}
\displaystyle
 \int_{-1}^{1} (k^2|\phi|^2+|\phi'|^2) dy &= \\
\displaystyle \int_{-1}^{1}
[f(y)\frac{(U-c_r)^2-(U_s-c_r)^2}{(U-c_r)^2+c_i^2}|\phi|^2]
dy&\leq
\\\displaystyle \int_{-1}^{1}
\frac{f(y)(U-c_r)^2}{(U-c_r)^2+c_i^2}|\phi|^2\, dy&.
\label{Eq:stable_paralleflow_Sun_Int_k2}
\end{array}
\end{equation}
Substituting $c_r=U_s$ and $\int_{-1}^{1} |\phi'|^2
dy>\mu_1\int_{-1}^{1}|\phi|^2 dy$ into it, this yields
\begin{equation}
k^2\int_{-1}^{1} |\phi|^2\, dy\leq   \int_{-1}^{1}
[\frac{f(y)(U-U_s)^2}{(U-U_s)^2+c_i^2}-\mu_1 ]|\phi|^2 dy.
\label{Eq:stable_paralleflow_Sun_Ineq_k_0}
\end{equation}
From the above inequality, we can obtain some new bounds for the
critical neutral stable wavenumber, the image of complex velocity,
and the growth rate.

First, a more strict upper bound for the critical neutral stable
wavenumber can be obtained by applying $\phi_1=\cos(\pi y/2)$ and
$c_i^2=0$ in inequality
(\ref{Eq:stable_paralleflow_Sun_Ineq_k_0}),
\begin{equation}
k^2\leq k_1^2=\int_{-1}^{1} [f(y)-\mu_1 ]\cos^2(\pi y/2) dy \leq
(p^2-1)\mu_1. \label{Eq:stable_paralleflow_Sun_Ineq_k1}
\end{equation}
The new upper bound $k_1^2$ of neutral stable wavenumber analogous
to the results in \cite{Craik1972}.

Second, we can also obtain the upper bound for $c_i^2$ from the
inequality (\ref{Eq:stable_paralleflow_Sun_Ineq_k_0}),
\begin{equation}
\begin{array}{rl}
\displaystyle   (\mu_1 + k^2)\int_{-1}^{1} |\phi|^2 dy  \leq &
\displaystyle \int_{-1}^{1}
\frac{f(y)}{1+c_i^2/(U-U_s)^2}|\phi|^2 dy \\
\leq & \displaystyle  \frac{1}{1+c_i^2/(\Delta U)^2} \int_{-1}^{1}
f(y) |\phi|^2 dy . \label{Eq:stable_paralleflow_Sun_Ineq_ci_0}
\end{array}
\end{equation}
where $\Delta U$ is the maximum of $U_{\max}-U_{s}$ and
$U_s-U_{\min}$. Thus by applying $\phi_1=\cos(\pi y/2)$ into
inequality (\ref{Eq:stable_paralleflow_Sun_Ineq_ci_0}), the upper
bound for $c_i^2$ is,
\begin{equation}
 c_i^2 \leq \frac{k_1^2-k^2}{\mu_1+k^2} \Delta U^2 \leq \frac{k_1^2}{\mu_1}\Delta U^2=c_{i0}^2
\label{Eq:stable_paralleflow_Sun_Ineq_ci0}
\end{equation}
It is obvious that $c_i^2\geq 0$ only when $k^2 \leq k^2_1$, which
covers the first result. This upper bound of $c_{i0}$ is much
better than that in \cite{Howard1961,Craik1972,Banerjee2000},
especially when the flow is slightly unstable.

Third, the upper bound for growth rate can be obtained by taking
$f(y)<p^2\mu_1$ in to inequality
(\ref{Eq:stable_paralleflow_Sun_Ineq_k_0}) and multiplying it by
$c_i^2$,
\begin{equation}
\omega_i^2\int_{-1}^{1} |\phi|^2\, dy\leq  \int_{-1}^{1}
h(y)|\phi|^2 dy, \label{Eq:stable_paralleflow_Sun_Ineq_hfunc1}
\end{equation}
where
\begin{equation}
h(y)=\mu_1[\frac{(p^2-1)(U-U_s)^2-c_i^2}{(U-U_s)^2+c_i^2}]c_i^2.
\label{Eq:stable_parallelflow_hfunc2}
\end{equation}
When $c_i^2=(p-1)(U-U_s)^2$, the right hand of
Eq.(\ref{Eq:stable_parallelflow_hfunc2}) get its largest value
\begin{equation}
h(y)= (p-1)^2 \mu_1 (U-U_s)^2.
 \label{Eq:stable_parallelflow_hy}
\end{equation}
Then the growth rate must be subject to
\begin{equation}
  \omega_i \leq (p-1)\sqrt{\mu_1} \Delta U,
  \label{Eq:stable_parallelflow_growth_omegamax}
\end{equation}
And the wavenumber $k_{m}$ corresponding to the largest growth
rate for $c_i^2=(p-1)(U-U_s)^2$ is approximately obtained by using
$k=\omega_i/c_i$,
\begin{equation}
  k_{m}\approx \sqrt{(p-1)\mu_1}.
  \label{Eq:stable_parallelflow_growth_max_k}
\end{equation}
So the results are proved. One should note that the fast growth
rate $\omega_i$ is only an approximate value, but not a precise
one, so as to the wavenumber $k_{m}$.

After we obtained the new bounds, we need to compare them with the
previous results. It is obvious that the present estimations are
stricter than the ones by H{\o}iland \cite{Hoiland1953} and Howard
\cite{Howard1961}. But we need still to compare them with the
results by Craik \cite{Craik1972} and Banerjee
\cite{Banerjee2000}.

First, compare Eq.(\ref{Eq:stable_paralleflow_Sun_Ineq_k1}) with
the studies by Craik \cite{Craik1972}. He used sinh functions to
express $\phi$, while we used the sin and cosine functions. As the
sine and cosine functions are orthogonal in the domain $a\leq y
\leq b$, our estimation should be more simple and better. Banerjee
et al. \cite{Banerjee2000} also noted this, they pointed out that
Tollmien's counter example \cite{Tollmien1935}, i.e., a sine or
cosine function, is very important. Both in \cite{Banerjee2000}
and in our previous work (Figure 2 in \cite{SunL2007ejp}), the
neutrally stable solution with $\phi_1=\cos(\pi y/2)$ was shown.

Second, compare inequality
(\ref{Eq:stable_paralleflow_Sun_Ineq_ci0}) with the results in
\cite{Craik1972,Banerjee2000}. In these studies, they simply
applied $c_i^2 \leq (U-U_s)^2+c_i^2$ into inequality
(\ref{Eq:stable_paralleflow_Sun_Ineq_k_0}). Thus the imaginary
part $c_i$ of the complex phase velocity is overestimation to $c_i
\leq \alpha/k^{p}$, where $\alpha$ and $p$ are two positive
constants. So $c_i$ is infinity as $k \rightarrow 0$, which can
also be seen from the figure 1 in \cite{Craik1972}. In this case,
the unbounded estimations are even looser than the Howard's
semicircle theorem \cite{Howard1961}. Craik \cite{Craik1972} also
used inequality $2(U-U_s)c_i \leq (U-U_s)^2+c_i^2$ into inequality
(\ref{Eq:stable_paralleflow_Sun_Ineq_k_0}) to obtain a better
estimation $c_i \leq \frac{1}{2}U''_{\max}/(k^2+\mu_1)$, which
analogous to and is stricter than H{\o}iland's result
\cite{Hoiland1953}. In contrast to that, we use a better
estimation in inequality
(\ref{Eq:stable_paralleflow_Sun_Ineq_ci0}). And $c_i$ has a
similar form of that in Howard's semicircle theorem
\cite{Howard1961}. When the flow is near unstable, i.e., $f(y)$
slightly bigger than $\mu_1$, the new estimation is much better
than all the previous ones.

Third, compare the new estimation of growth rate in
Eq.(\ref{Eq:stable_parallelflow_growth_omegamax}) with the results
in \cite{Craik1972,Banerjee2000}. Craik hardly obtained any useful
estimation \cite{Craik1972}, because his bound for $c_i$ is too
poor as $k \rightarrow 0$. Banerjee et al. \cite{Banerjee2000} got
a new one $\omega_i^2<(p^2-1)|U''|^2_{\max}/(p^2\mu_1)$, which is
better than H{\o}iland's \cite{Hoiland1953}. If we note that
$f(y)>\mu_1$, then $|U''|_{\max}>\mu_1 \Delta U$. Banerjee's
growth rate is approximately $\omega_i^2<(1-1/p^2)\mu_1 \Delta
U^2$, which is looser than the present bound in
Eq.(\ref{Eq:stable_parallelflow_growth_omegamax}). For example,
taking $p=1.1$, then $(p-1)^2=0.01\ll 0.17=(1-1/p^2)$.

A physical explanation on the long-wave instability and the K-H
instability would be that the K-H instability model has no
intrinsic length scale \cite{Huerre1998,CriminaleBook2003} . It
should be noted that Rayleigh's case is reduced to the
Kelvin-Helmholtz vortex sheet model under the long-wave limit
$k\ll 1$ \cite{Huerre1998,CriminaleBook2003}, which can be
explained as the long-wave not identifying with the finite
thickness of the shear layer \cite{Huerre1998}. In the present
study, we have shown that this explanation can be extended to
shear flows. Equation
(\ref{Eq:stable_parallelflow_growth_omegamax}) shows that the
growth rate $\omega_i$, is proportional to $\sqrt{\mu_1}$. Thus,
the thinner the shear layer, the larger the fastest wavenumber
becomes. The asymptotic case of the infinitely small shear layer
leads to K-H instability; this is another evidence that K-H
instability is essentially a long-wave instability. In this case,
K-H instability is an approximation of shear instability when the
length of the wave of perturbation is much longer than the width
of the shear layer.

The present studies also extend the previous theory on inverse
energy cascade in two dimensional flows \cite{Kraichnan1976}. When
studying the turbulence models, Kraichnan (1976) first found out
that the energy should transfer from small-scales (short-waves) to
large scales (long-waves), and that the eddy viscosity (for
small-scale eddies or short-wave perturbations) is negative. The
mechanism is interaction of large-scale straining fields with
small-scale vorticity fluctuations. Our previous study pointed out
that shear instability requires some conditions
\cite{SunL2008cpl}. First, a concentrated vortex is needed in the
flow. Second, the standing waves (with $c_r=U_s$) interact with
the concentrated vortex, so they can trigger instabilities. Third,
the waves must be large scale. The findings suggest that shear
instability in flows is due to the long-wave instability, and that
the energy should transfer to long-waves after instability. Our
present results extends the inverse energy cascade theory by
obtaining a fine estimation of wavenumber $k$ for large scale
motions, the maxi unstable wave phase speed ($c_r=U_s$), and a
better growth rate estimation for energy transfer calculation.

Besides, the present results are also very important for numerical
calculation. It implies that short-waves can be truncated in the
calculations without changing the stability of shear flow. In
contrast to that, the truncation of long-waves would probably
change the instability of the shear flow. So the streamwise length
scale must be longer enough to include long-waves for the
numerical simulations in shear flows, such as plane parallel flow
and pipe flow. Otherwise, the instability of shear flow would be
suppressed without long-wave perturbations.

In summary, three general properties of shear instability were
obtained in the investigation. First, short-waves were found to be
neutrally stable in the continuous profile flows, and that shear
instability was due to long-wave instability. We obtain a new
upper bound of $k_0$. Second, we find a new upper bound for the
imaginary part $c_i$ of the complex phase velocity. Third, we find
a new upper bound of growth rate $\omega_i \leq (p-1) \sqrt{\mu_1}
\Delta U$. All the new bounds are much more strict than the
previous ones. This estimate extends the previous results obtained
by H{\o}iland, Howard, Craik and Banerjee et al. Our results also
extend the inverse energy cascade theory by Kraichnan. These
findings are deemed useful for numerical calculations and
stability analysis.

We thanks the reviewers for bringing our notice the previous works
of \cite{Craik1972,Banerjee2000}. The useful comments by Huang
R.X. at WHOI and two anonymous reviewers are acknowledged. This
work is supported by the Knowledge Innovation Program of the
Chinese Academy of Sciences (Nos. KZCX2-YW-QN514), and the
National Basic Research Program of China (No. 2007CB816004).

\end{document}